\documentclass{PoS}
\usepackage{amsmath,amssymb}
\def\beq{\begin{equation}}
\def\eeq{\end{equation}}
\def\bea{\begin{eqnarray}}
\def\eea{\end{eqnarray}}
\def\BAN{\begin{eqnarray*}}
\def\EAN{\end{eqnarray*}}

\def\nn{\\ \nonumber}

\def\g5{\gamma_5}
\def\Id{\boldsymbol 1 }

\title{TWQCD's dynamical DWF project}

\ShortTitle{Lattice QCD with optimal domain-wall quarks}

\author{TWQCD Collaboration:
    \speaker{Ting-Wai~Chiu}$^{,1,2}$\thanks{Email: twchiu@phys.ntu.edu.tw}, 
    Tian-Shin~Guu$^{3}$, 
    Tung-Han~Hsieh$^{4}$,
    Chao-Hsi~Huang$^3$,
    Yu-Ying~Lee$^{1}$,  
    Yao-Yuan~Mao$^1$,
    Kenji~Ogawa$^1$, 
    Po-Kai~Tseng$^2$  \\
    $^1$ Department of Physics, and Center for Theoretical Sciences, National Taiwan University, Taipei 10617, Taiwan \\
    $^2$ Center for Quantum Science and Engineering, National Taiwan University, Taipei 10617, Taiwan \\
    $^3$ Center for General Education, and Institute of CSIE, National ILan University, I-Lan 260, Taiwan \\ 
    $^4$ Research Center for Applied Sciences, Academia Sinica, Taipei 115, Taiwan}

\abstract{
   We present an overview of our project of simulation of unquenched lattice QCD
   with optimal domain-wall quarks, 
   using a GPU cluster currently constituting of 16 units of Nvidia Tesla S1070 plus 
   64 graphic cards with Nvidia GTX285 (total 128 GPUs with 128 Teraflops peak),
   attaining sustained computing power of 15.36 Teraflops. 
   The first production run in two-flavor QCD is on-going, using the Iwasaki gauge action 
   on a set of lattices with sizes $ 16^3 \times (32,10,8,6,4) \times (16,32) $   
   at the lattice spacing $ a \sim 0.1$ fm, with eight sea quark masses down to $ m_\pi \simeq 200 $ MeV.  
   We outline our simulation algorithm, and describe the present status 
   of the production run. Preliminary results of pseudoscalar mass and decay constant are also presented. 
}

\FullConference{The XXVII International Symposium on Lattice Field Theory - LAT2009\\
		 July 26-31 2009\\
		 Peking University, Beijing, China}

\begin{document}

\section{Introduction}

Simulation of unquenched lattice QCD with exact chiral symmetry is a grand challenge
among all sciences. Even for a modest $ 16^3 \times 32 $ lattice with lattice 
spacing $ a \sim 0.1 $ fm, it often requires a supercomputer with peak computing power more than
50 Teraflops (e.g., 10 racks of IBM BlueGene/L). Therefore, only 2-3 lattice QCD 
groups around the world could afford to perform the simulation of unquenched lattice QCD with
the domain-wall fermion \cite{Kaplan:1992bt}, 
or the overlap-Dirac fermion \cite{Neuberger:1997fp}.
(For a recent review of generating gauge ensembles of QCD, see Ref. \cite{Jung:2009Abc}.)

However, this scenario has been undergoing a dramatic change during the last 9 months. 
With the emergence of low-cost and computationally powerful Graphic Processing Unit (GPU), 
now plugging a graphic card with Nvidia GTX285 (240 cores, one Teraflops peak) 
into a PC immediately turns the system into a powerful device, attaining a sustained 
140 Gflops for the conjugate gradient (CG) with mixed precision. 
This implies that one can perform the Hybrid Monte Carlo (HMC) simulation 
of unquenched lattice QCD with exact chiral symmetry, 
using a single PC with a GTX285 graphic card, rather than a cluster of 128 nodes. 
Moreover, the cost of such a system (PC + GPU card) could be less than \$600. 
In terms of price/performance, a GPU cluster is only 1/100 of a supercomputer (e.g., IBM BlueGene/L). 
Also, the power consumption of a GPU cluster is less than 1/10 of any supercomputer,  
for the same amount of (sustained) Teraflops.
Evidently, these salient features of GPU have revolutionary impacts to lattice QCD 
as well as all computational sciences. 
For a recent review of QCD on GPUs, see Ref. \cite{Clark:2009qp}.

The dynamical DWF project of TWQCD Collaboration was started around November 2007, 
soon after CUDA version 1.1 became available\footnote{
T.W.C. thanks Kelvin Chiu (then at Intel Corporation, CA) for his helpful suggestion.}. 
This was the right time for us to cut in  
since we could write our QCD code in CUDA rather than the much more tedious 
Cg \cite{Egri:2006zm}.
We completed our first CUDA code around March 2008, which implemented the 
nested conjugate gradient with mixed precision to compute the overlap fermion 
propagator, in which the inner CG (i.e., the sign function times a vector) 
was computed with the two pass algorithm \cite{Neuberger:1998jk, Chiu:2003ub}.
Note that Nvidia GTX280 had not been available at that time, 
thus all double precision operations were performed by the CPU, 
yet we still achieved 20 Gflops (sustained) using the  
Nvidia GTX8800 + Intel Q6600 (2.40 GHz). 

At lattice 2008, two groups reported using GPU for the Wilson-Dirac 
matrix multiplication with a vector \cite{Barros:2008rd,Ishikawa:2008pf},
attaining $\sim 100$ Gflops with Nvidia GTX280.
On the other hand, our aim had been to simulate unquenched lattice QCD
with exact chiral symmetry, in which the major difficulty 
is how to preserve chiral symmetry to a high precision 
and also to sample all topological sectors ergodically.  
For lattice QCD with the conventional domain-wall fermion \cite{Allton:2007hx}, 
it suffers from a large chiral symmetry breaking (i.e., large residual mass) 
for $ N_s = 16 $, which becomes problematic for finite temperature QCD. 
On the other hand, for lattice QCD with overlap-Dirac fermion,  
the action is discontinuous at the boundary between different topological sectors, 
and it is very costly to tunnel through the topological boundary using the 
refraction/reflection algorithm \cite{Fodor:2003bh}.
A plausible way to proceed is to fix the topology \cite{Fukaya:2006vs}, 
then use the measured values of topological susceptibility to remove 
the artifacts in any physical observables due to the fixed topology 
in a finite volume. However, it has not been proved that suppressing the near zero modes
(with the unphysical Wilson fermions and the associated twisted-mass ghosts)  
does not violate the ergodicity in a fixed topological sector. 

Now we turn to the optimal domain-wall fermion. 
Unlike the conventional domain-wall fermion  
suffering from large chiral symmetry breaking, the optimal domain-wall fermion   
attains the (mathematically) maximal chiral symmetry by judiciously fixing the 
weight $ \omega_s $ for each layer in the fifth dimension,  
according to the formula derived in Ref. \cite{Chiu:2002ir}.   
Then the effctive 4D Dirac operator is exactly equal to the overlap-Dirac 
operator with the sign function approximated by the Zolotarev optimal rational polynomial. 

The first step of our project was to develop the CUDA code for the CG 
of the 5D optimal domain-wall quark matrix, the most computationally intensive part in HMC. 
After tuning our CUDA code for 5 months, in January 2009,  
we achieved 120 Gflops for a single GTX280. 

The outline of this paper is as follows. 
In Section 2, we describe the hardware of our project. 
In Section 3, we outline our HMC simulation with the optimal domain-wall fermion.
In Section 4, we briefly describe the GPU optimization of our CG code. 
In Section 5, we describe the present status of our production runs. 
In Section 6, we present preliminary results of pseudoscalar mass and decay constant.
Finally, in Section 7, we conclude with some remarks.

\section{The hardware}

Our hardware is a GPU cluster currently constituting of 16 units of Nvidia Tesla S1070 
plus 64 Nvidia GTX285 graphic cards. 
Each unit of Nvidia Tesla S1070 contains 4 Tesla GPUs, 
each having 4 Gbyte RAM, while each GTX285 has 1 or 2 Gbyte RAM.    
Each unit of Tesla S1070 is connected to the PCI Express x16 slots of 
a Tyan server through two interconnect cables and two interface cards.  
Each Tyan Server has two Intel Quad-Core Xeon processors and 32 GB RAM. 
For the GTX285 cards, every two GTX285 cards are plugged into the PCI Express x16 slots of a 2U server 
with one Intel i7 CPU and 12 Gbyte RAM. 
The hard disk storage of our system is over 300 Terabytes, which is managed  
through the Lustre cluster file system (http://www.sun.com/software/products/lustre/index.xml).
The hardware cost of our system is around \$220,000.
The sustained computing power of our system attains 15.36 Teraflops 
for HMC simulation of lattice QCD with optimal domain-wall quarks.
We use the Condor to manage the workload of our computing jobs (see http://www.cs.wisc.edu/condor/).

\section{Hybrid Monte Carlo simulation with optimal domain-wall quark}

The action of optimal domain-wall fermion \cite{Chiu:2002ir} is defined as   
\bea
\label{eq:ODWF}
S_{odwf} 
&=& \sum_{s,s'=1}^{N_s} \sum_{x,x'} \bar\psi_{xs}
\{ (\omega_s D_w + \Id )_{xx'} \delta_{ss'}  + (\omega_s D_w - \Id )_{xx'}
    (P_{+} \delta_{s',s-1} + P_{-} \delta_{s',s+1} ) \} \psi_{x's'} \nn
&=& \bar\Psi D_{odwf} \Psi 
\eea
with boundary conditions
$ P_{+} \psi_{x,0} = - (m_q/2 m_0) P_{+} \psi_{x,{N_s}} $, 
$ P_{-} \psi_{x,{N_{s+1}}} = - (m_q/2 m_0) P_{-} \psi_{x,1} $, where $ P_{\pm} = (1 \pm \gamma_5)/2 $. 
Here $ D_w $ is the standard Wilson Dirac
operator plus a negative parameter $ -m_0 $ ($ 0 < m_0 < 2 $), 
$ m_q $ is the bare quark mass,  
and the weights $ \{ \omega_s \} $ are specified  
by the exact formula derived in Ref. \cite{Chiu:2002ir}.
The optimal domain-wall fermion operator $ D_{odwf} $ can be transformed to
\bea
\label{eq:Dopt}
D_{opt} = D_w + M, \hspace{4mm} 
M = \frac{1}{\sqrt{\boldsymbol \omega}} \frac{\Id-L}{\Id+L} \frac{1}{\sqrt{\boldsymbol \omega}},  
\hspace{4mm}
L = P_+  L_+ + P_- L_- , 
\eea
where the $\boldsymbol \omega $ denotes the diagonal matrix of the weights $ \{ \omega_s, s=1, \cdots, N_s \} $, 
and 
\beq
{(L_+)}_{s,s'} =
\begin{cases}
 \delta_{s',s - 1},  &  1 < s \le N_s,    \\
 -(m_q/2 m_0) \delta_{s', N_s}, &  s = 1, 
\end{cases}
, \hspace{6mm}
L_- = (L_+)^{\dagger} .
\eeq
Obviously, $ M $ does not carry the color and 4D space-time indices, and it can be solved analytically. 
Since $ \det(D_{opt}) $ only differs from $ \det(D_{odwf}) $ by a factor independent of the gauge field, 
we can use $ D_{opt} $ to perform the HMC simulation rather than $ D_{odwf} $.     

With even-odd preconditioning in the 4D space-time lattice, $ D_{opt} $ can be written as 
\bea
\label{eq:Dopt_eo}
D_{opt} =
\begin{pmatrix}
X &  D_w^{eo}  \\
D_w^{oe}    & X 
\end{pmatrix}
=
\begin{pmatrix}
\Id &  0  \\
D_w^{oe} X^{-1}   & \Id 
\end{pmatrix}
\begin{pmatrix}
X &  0  \\
0 & X - D_w^{oe} X^{-1} D_w^{eo}  
\end{pmatrix}
\begin{pmatrix}
\Id &  X^{-1} D_w^{eo}  \\
0   & \Id 
\end{pmatrix}
\eea
where $ X = 4-m_0+M $. Since $ \det D_{opt} = \det( 1 - D_w^{oe} X^{-1} D_w^{eo} X^{-1} ) \det X $, and 
$ X $ does not depend on the gauge field, we can use the operator 
\bea
\label{eq:C}
C = 1 - D_w^{oe} X^{-1} D_w^{eo} X^{-1}  
\eea 
for the HMC simulation. After including the Pauli-Villars fields (with $ m_q = 2 m_0 $), the pseudo-fermion 
action for 2-flavor QCD ($ m_u = m_d $) can be written as 
\bea
S_{pf} = \phi^\dagger C_{PV}^\dagger ( C C^\dagger)^{-1} C_{PV} \phi , 
\eea
where $ C_{PV} = C(m_q=2m_0) $. 

In the HMC simulation, we first generate random noise vector $ \xi $ with Gaussian distribution,  
then we obtain $ \phi = C_{PV}^{-1} C \xi \equiv D \xi $ with CG.
With fixed $ \phi $, the system is evolved with a fictituous Hamiltonian dynamics, 
the so-called molecular dynamics (MD). In the MD, we use the Omelyan integrator \cite{Takaishi:2005tz},
and the Sexton-Weingarten multiple-time scale method \cite{Sexton:1992nu}. 
The most time-consuming part in the MD is to compute $ \eta = (D D^\dagger)^{-1} \phi $ 
with CG, which is required for the evaluation of the fermion force in the equation
of motion of the conjugate momentum of the gauge field. Thus we program the GPU (Nvidia T10 or GTX285) 
to compute $ \eta = (D D^\dagger)^{-1} \phi $, using CG with mixed precision. 
Also, we have ported the computation of the gauge force and the gauge field update to the GPU. 
Our next step is to implement the computation of the fermion force entirely inside GPU. 
In some of our HMC simulations, we also introduce an extra heavy fermion field to reduce
the condition number of the CG involving the light quark \cite{Hasenbusch:2001ne}. 
In this case, we can have both CPU and GPU compute concurrently,  
e.g., while the GPU is working on the CG of the light quark field, 
the CPU can compute the fermion force of the heavy fermion field. 
This asynchronous concurrent excecution mode enhances the overall performance by $\sim5$\%.

\section{GPU optimizations}

The Nvidia CUDA programming guide (see http://developer.download.nvidia.com) has provided 
general guidelines to optimize an application code for CUDA. 
Here we outline some useful techniques in tuning our CUDA code 
for CG (with mixed precision) of the (optimal) domain-wall quark matrix. 
First, we have one thread take care of all computations at each $ (s,x,y,z) $, 
with a loop going over all $t$ (even/odd). The threads are divided into a one-dimensional grid with 
two-dimensional blocks. Each block has 64 threads, with $ N_s $ threads in the X-direction and 
$ 64/N_s $ threads in the Y-direction. The number of blocks is equal to 
$ N_s N_x N_y N_z /64 $. 
At first, a naive transcription of our C code to CUDA 
only yielded 15 Gflops on a single GTX280. After putting the constant matrix $ X^{-1} $ 
(only involving the 5-th dimensional and Dirac indices) in the constant memory space, 
and using the texture memory for link variables and vectors, we obtained an enhancement  
from 15 to 30 Gflops. Then we unrolled the short loops 
(involving Dirac and color indices) in our CG code. 
This gave a further enhancement from 30 to 50 Gflops.     
Next we reordered the data in the global memory such that the memory access by all threads
of a half-wrap is coalesced into one memory transaction. This enhanced the performance
from 50 to 80 Gflops. Then we reduced the number of local variables in our code, which
gave another jump from 80 to 100 Gflops. Finally, we combined the threads
by re-using forward/backward data (in $t$) for neighboring sites.   
This increased the performance from 100 to 120 Gflops.
For Nvidia T10/GTX285, our CG code attains 100/140 Gflops,  
where the difference is mainly due to the memory bandwidth and clock frequency 
of the GPU. For the T10 processor with 4 GB device memory, it can only accommodate a lattice up to 
the size $ 24^3 \times 48 \times 16 $. In order to simulate a larger lattice,  
we are porting our CG CUDA code to multi-GPUs under MPI and OpenMP.

\section{Current status of production runs}

\begin{figure}[tb]
\begin{center}
\begin{tabular}{@{}c@{}c@{}}
\includegraphics*[height=5.8cm,width=7.55cm]{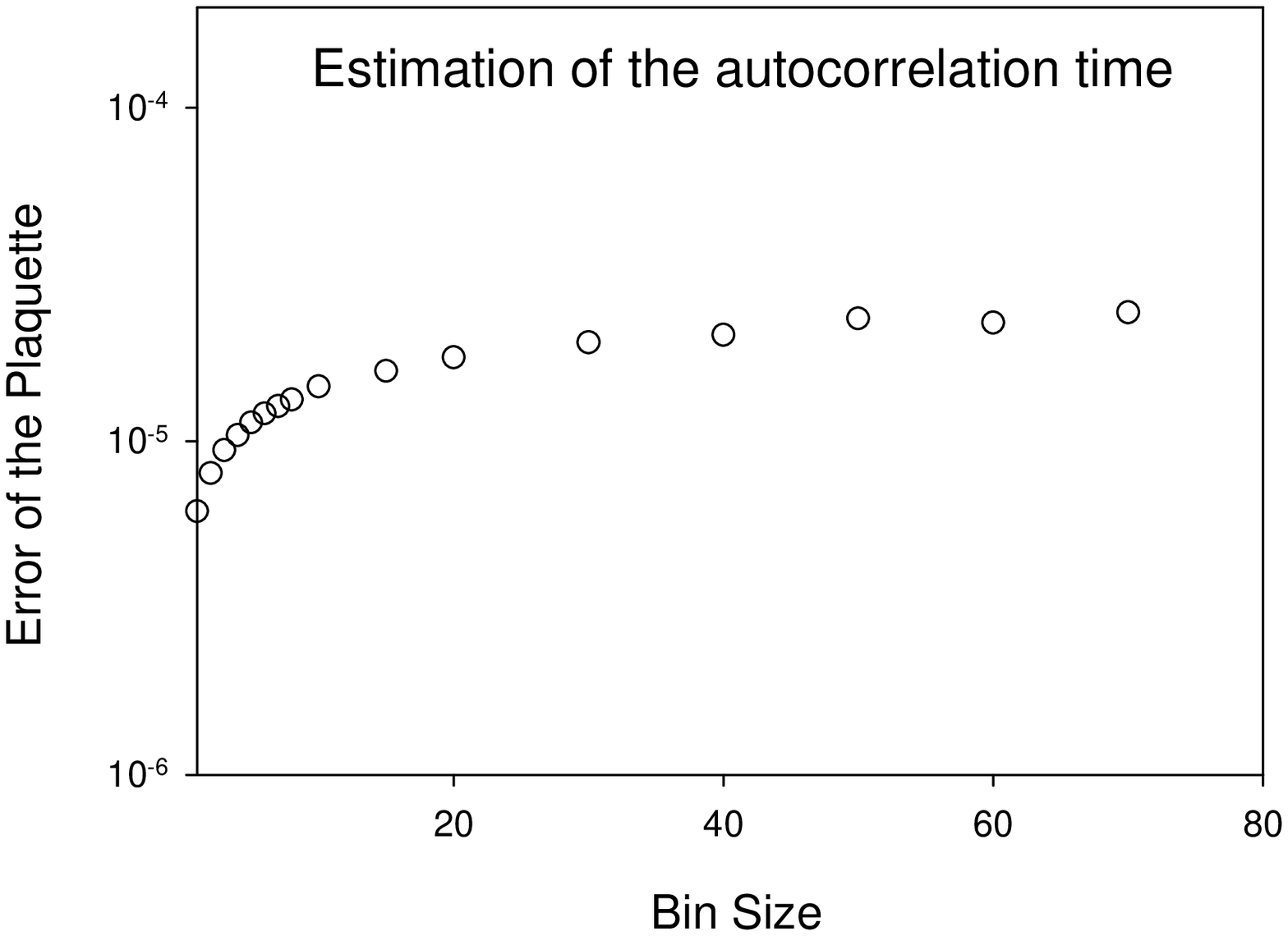}
&
\includegraphics*[height=5.8cm,width=7.55cm]{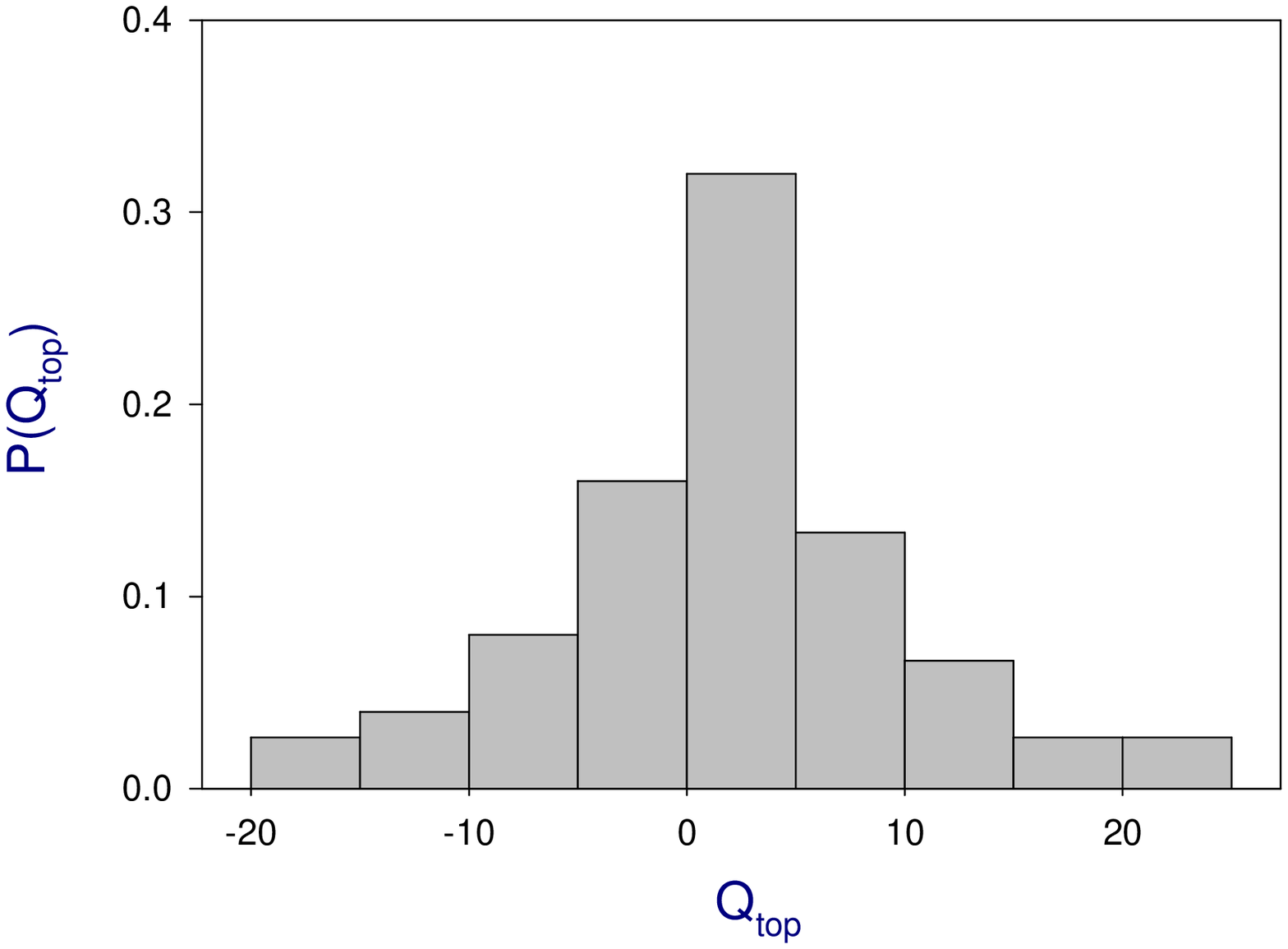}
\\ (a) & (b)
\end{tabular}
\caption{(a) Estimation of the autocorrelation time using the binning method. 
         The jackknife error of the plaquette is measured from 4000 accepted
         trajectories in the HMC simulation of 2-flavor QCD 
         with optimal domain-wall quarks ($ m_q a = 0.02 $) 
         on the $ 16^3 \times 32 \times 16 $ lattice. 
         (b) The topological charge distribution of 66 configurations which are 
         sampled every 20 accepted trajectories.}
\label{fig:auto}
\end{center}
\end{figure}

The first production run in two-flavor QCD is on-going, using the Iwasaki gauge action 
on a set of lattices with sizes $ 16^3 \times (32,10,8,6,4) \times (16, 32) $   
at the lattice spacing $ a \sim 0.1$ fm ($ \beta = 2.20 $), 
with eight sea quark masses ranging from $ m_q a = 0.01, 0.02, \cdots, 0.08 $.  
After discarding 300 trajectories for thermalization, we are accumulating 
$ 5120 $ trajectories in total for each sea quark mass.
Using the binning method, we estimate the autocorrelation time to be around 20,  
based on the saturation of the jackknife error of the plaquette 
as a function of the bin size, as shown in Fig. \ref{fig:auto}(a).  
Thus we sample one configuration every $ 20 $ accepted trajectories, 
and we have $ 256 $ configurations for each $ m_q $. 
The topological charge distribution of 66 configurations for $ m_q = 0.02 $ 
is plotted in Fig. \ref{fig:auto}(b).  
Currently, we are completing the statistics for each gauge ensemble, 
and we expect to have the first physics results from the ensemble with the 
lattice size $ 16^3 \times 32 \times 16 $.

\section{Preliminary results of pseudoscalar mass and decay constant}

\begin{figure}[tb]
\begin{center}
\begin{tabular}{@{}c@{}c@{}}
\includegraphics*[height=5.8cm,width=7.55cm]{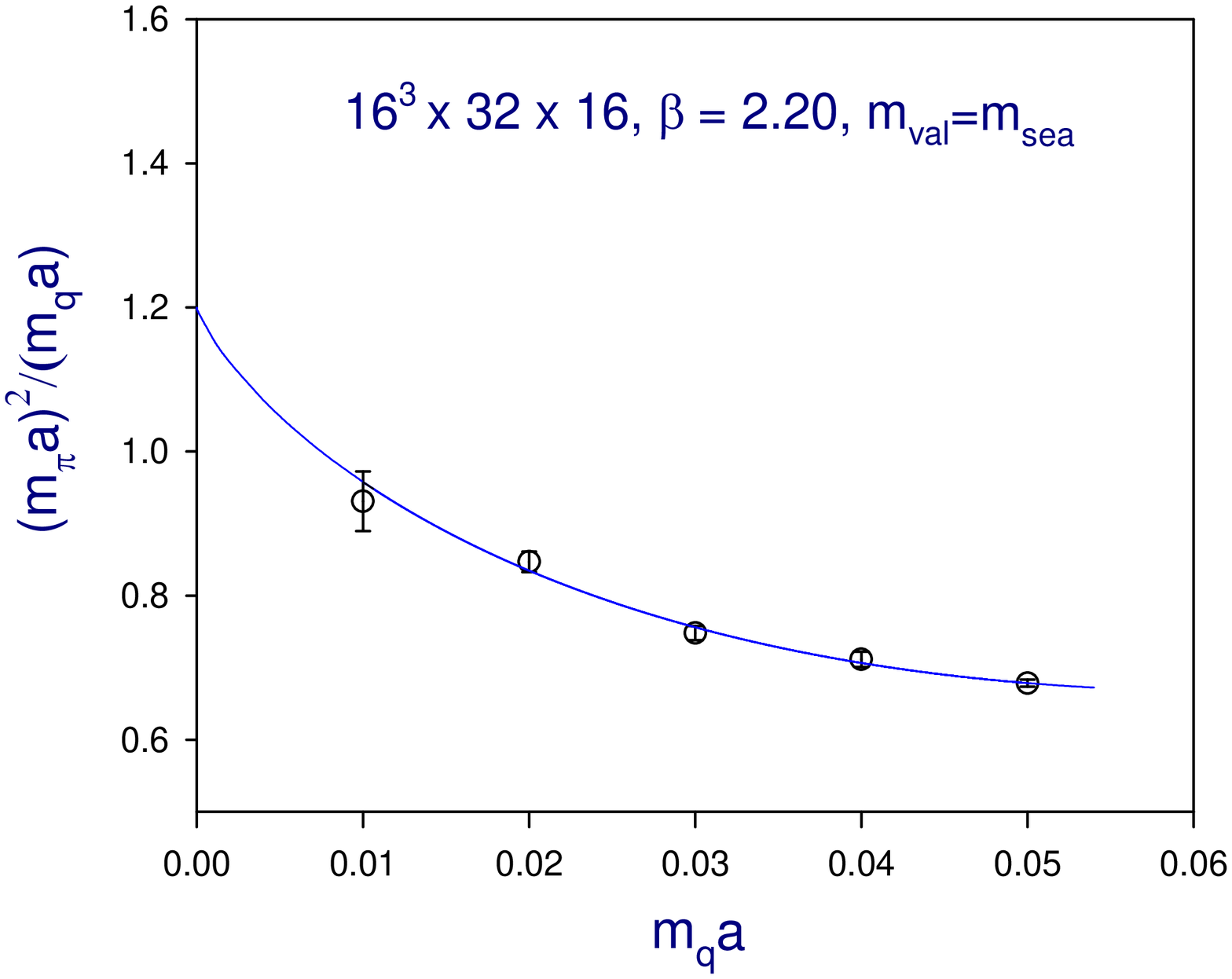}
&
\includegraphics*[height=5.8cm,width=7.55cm]{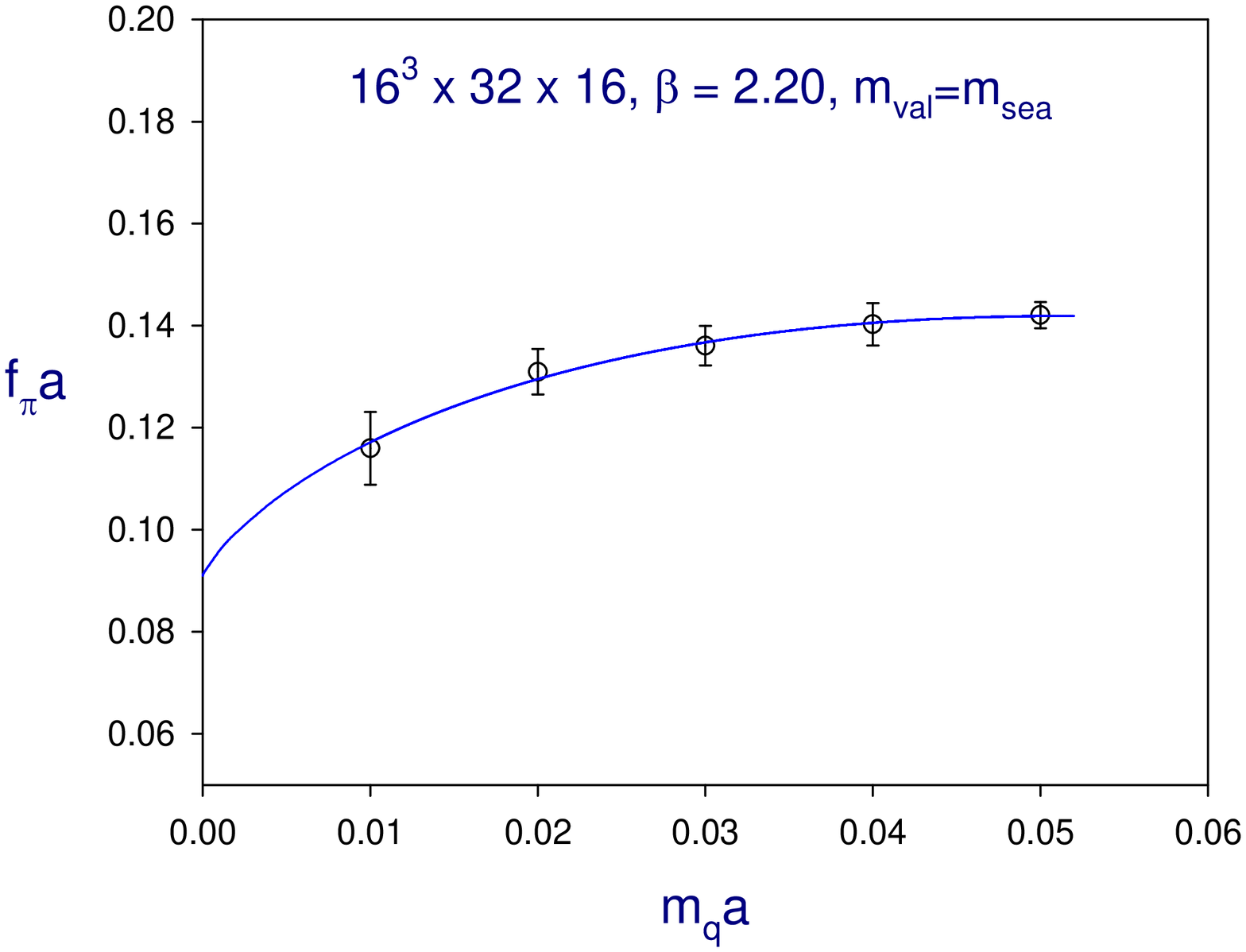}
\\ (a) & (b)
\end{tabular}
\caption{ Preliminary results of 2-flavor QCD with optimal domain-wall quarks: 
          (a) $ m_\pi^2/m_q $, and (b) $ f_\pi $.
          The solid lines are the fits to the ChPT at NLO.} 
\label{fig:mpi2omq_fpi_b220_nf2}
\end{center}
\end{figure}

In this section, we present some preliminary results of the pseudoscalar mass and decay constant, 
for the gauge configurations generated with 2-flavor QCD with optimal domain-wall quarks 
and the Iwasaki gluon action at $ \beta = 2.20 $, on the $ 16^3 \times 32 \times 16 $ lattice. 
The weights $ \{ \omega_s \} $ are fixed such that the error of the sign function $ S_{opt}(H_w) $ 
is less than $ 10^{-5} $.
In Fig. \ref{fig:mpi2omq_fpi_b220_nf2}, we plot $ m_\pi^2 /m_q $ and $ f_\pi $ versus $ m_q $
respectively. The statistics for each quark mass is about 100-150 configurations.
We fit our results to the next-to-leading order (NLO) formulas 
in chiral perturbation theory (ChPT) \cite{Gasser:1984gg}
\bea
\label{eq:mpi2omq_NLO}
\frac{m_\pi^2}{m_q} &=& 2 B \left( 1 + \frac{1}{2} x \ln x \right) + c_3 x,  \hspace{6mm} 
x = \frac{4 B}{(4 \pi f)^2} m_q,  \\
\label{eq:fpi_NLO}
f_\pi &=& f ( 1 - x \ln x ) + c_4 x .
\eea 
Using the fitted value of $ f a $ and the physical input $ f_\pi = 131 $ MeV, we obtain
$ a^{-1} = 1.5(1) $ GeV, a rough estimate of the lattice spacing. 
From the fitted value of $ B $, we obtain the chiral 
condensate $ \Sigma = B f^2 /2 = [261(20) \mbox{MeV}]^3 $.
The complete analysis with the statistics of 250 configurations 
for each quark mass will be presented in a forthcoming paper.

\section{Concluding remark}

This is the first project to use a GPU cluster to perform a large-scale simulation 
of lattice QCD with exact chiral symmetry. Our GPU cluster is   
less than 1/100 in price/performance ratio and only 1/10 of the power consumption 
of a conventional supercomputer.
Using the optimal domain-wall fermion, we can simulate unquenched lattice QCD with exact chiral symmetry,
without fixing the topology. The cost of preserving chiral symmetry to a high precision 
is counterbalanced by the low price/performance ratio of a GPU. 
The production runs for 2-flavor QCD on the $ 16^3 \times 32 \times 16 $ lattice 
will be completed by the end of 2009.     
For the (2+1)-flavor QCD on the $ 16^3 \times 32 \times 16 $ lattice, 
we simulate the strange quark with our newly constructed positive-definite pseudofermion action 
for a single fermion flavor \cite{Ogawa:2009ex}. The thermalization of this set of gauge ensembles 
is now in progress.

\begin{acknowledgments}
  This work is supported in part by  
  the National Science Council 
  (Nos. NSC96-2112-M-002-020-MY3, 
        NSC96-2112-M-001-017-MY3, 
        NSC98-2119-M-002-001),    
  and NTU-CQSE~(Nos. 98R0066-65, 98R0066-69).
  T.W.C. also thanks Nvidia Corporation for the support under Professor Partnership Program.  
\end{acknowledgments}

\end{document}